\documentstyle[]{mn}
\newif\ifAMStwofonts

\ifoldfss
  \ifCUPmtlplainloaded \else
    \NewTextAlphabet{textbfit} {cmbxti10} {}
    \NewTextAlphabet{textbfss} {cmssbx10} {}
    \NewMathAlphabet{mathbfit} {cmbxti10} {} 
    \NewMathAlphabet{mathbfss} {cmssbx10} {} 
  \fi
  \ifAMStwofonts
    \ifCUPmtlplainloaded \else
      \NewSymbolFont{upmath} {eurm10}
      \NewSymbolFont{AMSa} {msam10}
      \NewMathSymbol{\upi}     {0}{upmath}{19}
      \NewMathSymbol{\umu}     {0}{upmath}{16}
      \NewMathSymbol{\upartial}{0}{upmath}{40}
      \NewMathSymbol{\leqslant}{3}{AMSa}{36}
      \NewMathSymbol{\geqslant}{3}{AMSa}{3E}

      \let\leq=\leqslant 
      \let\geq=\geqslant \let\ge=\geqslant
    \fi
  \fi
\fi 

\ifnfssone
  \newmathalphabet{\mathit}
  \addtoversion{normal}{\mathit}{cmr}{m}{it}
  \addtoversion{bold}{\mathit}{cmr}{bx}{it}
  \newmathalphabet{\mathbfit} 
  \addtoversion{normal}{\mathbfit}{cmr}{bx}{it}
  \addtoversion{bold}{\mathbfit}{cmr}{bx}{it}
  \newmathalphabet{\mathbfss} 
  \addtoversion{normal}{\mathbfss}{cmss}{bx}{n}
  \addtoversion{bold}{\mathbfss}{cmss}{bx}{n}
  \ifAMStwofonts
    \ifCUPmtlplainloaded \else
      \UseAMStwoboldmath
      \makeatletter
      \new@mathgroup\upmath@group
      \define@mathgroup\mv@normal\upmath@group{eur}{m}{n}
      \define@mathgroup\mv@bold\upmath@group{eur}{b}{n}
      \edef\UPM{\hexnumber\upmath@group}
      \new@mathgroup\amsa@group
      \define@mathgroup\mv@normal\amsa@group{msa}{m}{n}
      \define@mathgroup\mv@bold\amsa@group{msa}{m}{n}
      \edef\AMSa{\hexnumber\amsa@group}
      \makeatother
      \mathchardef\upi="0\UPM19
      \mathchardef\umu="0\UPM16
      \mathchardef\upartial="0\UPM40
      \mathchardef\leqslant="3\AMSa36
      \mathchardef\geqslant="3\AMSa3E

      \let\leq=\leqslant 
      \let\geq=\geqslant \let\ge=\geqslant
    \fi
  \fi
\fi 

\ifnfsstwo
  \DeclareMathAlphabet{\mathbfit}{OT1}{cmr}{bx}{it}
  \SetMathAlphabet\mathbfit{bold}{OT1}{cmr}{bx}{it}
  \DeclareMathAlphabet{\mathbfss}{OT1}{cmss}{bx}{n}
  \SetMathAlphabet\mathbfss{bold}{OT1}{cmss}{bx}{n}
  \ifAMStwofonts
    \ifCUPmtlplainloaded \else
      \DeclareSymbolFont{UPM}{U}{eur}{m}{n}
      \SetSymbolFont{UPM}{bold}{U}{eur}{b}{n}
      \DeclareSymbolFont{AMSa}{U}{msa}{m}{n}
      \DeclareMathSymbol{\upi}{0}{UPM}{"19}
      \DeclareMathSymbol{\umu}{0}{UPM}{"16}
      \DeclareMathSymbol{\upartial}{0}{UPM}{"40}
      \DeclareMathSymbol{\leqslant}{3}{AMSa}{"36}
      \DeclareMathSymbol{\geqslant}{3}{AMSa}{"3E}

      \let\leq=\leqslant 
      \let\geq=\geqslant \let\ge=\geqslant
    \fi
  \fi
\fi 

\ifCUPmtlplainloaded \else
  \ifAMStwofonts \else 
    \def\upi{\pi}
    \def\umu{\mu}
    \def\upartial{\partial}
  \fi
\fi

\title[A metal poor HII galaxy at $z$=3.36]
{Nebular and stellar properties of a metal poor 
HII galaxy at $z=$3.36.}
\author[Villar-Mart\'\i n, Cervi\~no \& Gonz\'alez Delgado]
{M. Villar-Mart\'\i n$^1$\thanks{e-mail: montse@iaa.es}, M. Cervi\~no$^1$ \& R.M. Gonz\'alez Delgado$^1$\\
$^1$Instituto de Astrof\'\i sica de Andaluc\'\i a (CSIC), Aptdo. 3004, 18080,Granada, Spain\\}

\date{}

\pagerange{\pageref{firstpage}--\pageref{lastpage}}
\pubyear{2001}

\begin{document}

\maketitle

\label{firstpage}

\begin{abstract}

We have characterized the physical properties (electron temperature,
density, metallicity) of the ionized gas and
the ionizing population (age, metallicity, presence of WR stars) in
the Lynx arc, a HII galaxy  at $z=$3.36. The UV doublets (CIII], SiIII] and NIV) imply the
existence of a density gradient in this object, with a high density 
region (0.1-1.0 $\times$ 10$^5$ cm$^{-3}$) and a lower density region
($<$3200  cm$^{-3}$). The temperature sensitive ratio [OIII]$\lambda\lambda$1661,1666/$\lambda$5007
implies an electron temperature 
$T_e$=17300$^{+500}_{-700}$ K, in agreement within the errors with photoionization model predictions.  Nebular abundance determination
using standard techniques 
 and  the results from photoionization models imply a nebular metallicity of O/H$\sim$10$\pm$3\% (O/H)$_{\odot}$, in good agreement with 
Fosbury et al (2003).
Both methods suggest that nitrogen is overabundant relative to other
 elements, with [N/O]$\sim$2.0-3.0 $\times$ [N/O]$_{\odot}$.
 We do not find evidence for
Si overabundance, as Fosbury et al. (2003). 

Photoionization models imply that the ionizing stellar population in the Lynx arc
has an age of $\la$5 Myr. If He$^{+}$ is ionized by WR stars, then the ionizing
stars in  the Lynx arc have  metallicities Z$_{star}$$>$5\% Z$_{\odot}$ and 
ages $\sim$2.8-3.4 Myr (depending on Z$_{star}$), when WR stars appear and 
are responsible for the He$^{+2}$ emission.
However, alternative excitation mechanisms for this species are not discarded.
Since the emission lines trace the properties of 
the present burst only, nothing can be said
about the possible presence of an underlying old stellar population.

The Lynx arc is  a low metallicity HII galaxy  that
is undergoing a  burst of star formation of $\la$5 Myr  age.  One possible scenario that
 explains the emission line spectrum of the Lynx arc,
the large strength of the nitrogen lines and the 
He$^{+2}$ emission is that the object has experienced a merger event that has
triggered a burst of star formation.
 WR stars have formed that contribute to a fast enrichment of the ISM. 

As Fosbury et al. (2003), we find a factor of $\ga$10 discrepancy between the mass  of the 
instantaneous burst required to power  the luminosity of the H$\beta$ line and the mass
implied by the continuum level
measured for  the Lynx arc. We discuss several possible solutions to this problem.
The most likely explanation is that gas and stars have different spatial distributions
so that the emission lines and the stellar continuum suffer different gravitational 
amplification by the intervening cluster.

\end{abstract}


\section{Introduction}

H II galaxies are dwarf emission-line galaxies undergoing a burst of
star formation. They are characterized by strong and narrow emission 
lines originated in a giant star-forming region which dominate
their observable properties at optical wavelengths (e.g. \cite{ter04}). Most are blue compact galaxies (BCGs). 
They  have very low metallicities, high rates of star formation and a very young stellar
content. Many are compact and isolated.
One of the reasons why these objects have  attracted
 significant  attention
is the possibility that they are very young galaxies in the process
of formation.
This possibility, however, has been challenged
since 
  evidence for an old  (several Gyr) stellar
population has been found in numerous BCGs 
(see  \cite{kunth00} for a review). Therefore, a model with a 
single, instantaneous burst of star formation does not seem appropriate to 
describe these galaxies and a succession of short starbursts separated
by quiescent periods seems more likely (\cite{ter04}).

Even in the case if HII galaxies are not primeval, little chemical
evolution has happened and  they
  provide important information about how galaxies form and
evolve, as well as about the process of star formation in
low metallicity environments.

The most distant HII galaxy in the catalogue  compiled by 
Terlevich et al. (1991) is at redshift $z\sim$0.31, while the vast majority
are at $z<$0.1. The proximity of these galaxies allows studies of their
structure, metal content and stellar population with high sensitivity
and precision, difficult to achieve for high redshift star forming galaxies.
Still, finding high redshift HII galaxies is of great interest to
investigate their abundance patterns and stellar properties
 at earlier
epochs,  compare them with the nearby counterparts and obtain further information about the formation  
and evolution of galaxies using objects
 that have undergone little chemical evolution.

The Lynx arc is a star forming object at z$=$3.36. It was discovered
during spectroscopic follow-up of the  cluster 
 RXJ 0848+4456 (z=0.57) from the $ROSAT$ Deep Cluster Survey 
 (\cite{hol01}). The arc  shows a very red R-K color and very strong
optical and UV (rest frame) 
narrow emission lines.  From the analysis of HST WFPC2 images and Keck
optical and near infrared spectroscopy,
Fosbury et al. (2003, FOSB03 hereafter) concluded that the arc is an   HII galaxy
 magnified by
a factor of $\sim$10 by a complex intervening cluster environment. The authors concluded
that the continuum is mostly nebular.

By means of photoionization modeling,  \cite{fos03}  showed 
 that the spectroscopic properties of the Lynx arc are all consistent
with a simple HII region model of a cluster of $\sim$10$^6$ massive stars,
 characterized by a very high T$_{eff}\sim$80 000 K, a
high ionizing parameter (U$\sim$0.1) and  low nebular metallicity
Z$\sim$ 5\% Z$_{\odot}$. Indirect arguments suggest
that the ionizing stellar population could have much lower 
abundances. Using simple photoionizaton models,
the authors conclude that the  spectroscopic properties of the
Lynx arc are consistent with those of a metal-poor nebula ionized by a
cluster of  primordial (i.e. metal-free) stars.
In this scenario, the overabundance 
of Si  implied by the models is explained  as due 
to enrichment by past pair-instability supernovae, requiring
stars more massive than 120 M$_{\odot}$.

We investigate in this paper the possibility that normal (i.e. non primordial)
stars are responsible for the excitation of the gas in the Lynx arc.
Our first  goal is to constrain the physical properties of the gas: electron temperature,
density, metallicity. We will first use
 standard spectroscopic techniques
of nebular analysis (e.g. \cite{aller84}). These results will be
compared with those from detailed photoionization modeling and 
those obtained by \cite{fos03}.

 The second  goal is to characterize the  ionizing 
stellar population: age, metallicity, stellar mass of the burst,
possible presence of WR stars. 
This will be achieved by means of detailed photoionization models, whose 
objective is
to find an ionizing stellar population able to explain all the spectroscopic
properties of the Lynx arc (line ratios, equivalent widths, intensity of
the continuum). In addition,  such models set tight constrains on the nebular
properties, which must be consistent with the results from the standard techniques
mentioned above.

We present in \S2 the results of the nebular
 analysis and the photoionization models.
These results will be discussed in \S3. Summary and conclusions are presented in \S4.

\section{Analysis}

\subsection{Nebular properties}

In this section we will set constrains on the electron temperature, density  and the nebular 
abundances of the
ionized gas in the Lynx arc. 

All the analysis and discussion presented below are based on spectra obtained
 with different slit widths and orientations (see \cite{fos03} for
a detailed description of the data set and reduction techniques). For this reason, \cite{fos03}  estimated carefully scaling factors that were applied in order to derive
a consistent calibration. We will therefore assume that the errors introduced by
calibration uncertainties are negligible. 

The results are also dependent on whether reddening is present or not.
We have ignored this effect.  The very
strong UV lines and the lack of evidence for dust reddening from the photoionization
models and the fit to the continuum shape (\cite{fos03}) makes as confident that such assumption is
reasonable and our results are not seriously affected by it.  Our own
work presented in this paper will show that
the line ratios are consistent with no dust reddening.

\subsubsection{Electron temperature}

The  [OIII]4363/[OIII]5007 ratio has been traditionally used to
measure electron temperatures in ionized nebulae. Unfortunately,
the [OIII]$\lambda$4363
line is outside the observed spectral range. We have used the
UV [OIII] doublet instead, since 
the [OIII]$\lambda\lambda$1661,1666/[OIII]$\lambda$5007 ratio is mostly sensitive to electron temperature.

We show in Fig.1 the dependence of the [OIII]$\lambda\lambda$1661,1666/[OIII]$\lambda$5007 ratio with
electron temperature 
The position of the Lynx arc (the value of the ratio is 0.074$\pm$0.005, see Table 1) is also shown as a filled
circle. It implies  $T_e\sim$17300$^{+500}_{-700}$ K.
 This value is somewhat higher
than that predicted  by the photoionization models presented in $\S$2.2
which give $T_e=$16200$\pm$500 K, although taking errors into account, the discrepancy
is not important.
This further supports that dust reddening effects are negligible, otherwise,
the temperature derived from Fig. 1  would be a lower limit and 
large discrepancies with the photoionization model predictions would arise.
 This temperature is much lower than that derived by \cite{fos03}  and \cite{bin03}
by means of photoionization modeling ($\sim$20000 K). The discrepancy is due to a combination of
the very hard ionizing continuum (very  energetic electrons are therefore released in
the ionization processes) and high $U$ (ionization parameter) value used by these authors.

\begin{figure}
\includegraphics{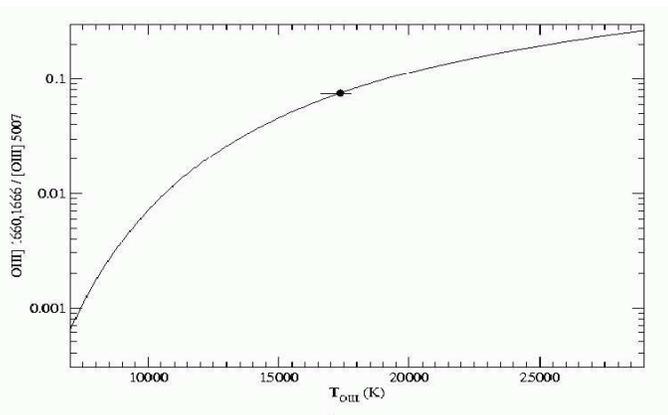}
\vspace{2.5in}
\caption{Dependence of the [OIII]$\lambda\lambda$1661,1666/[OIII]$\lambda$5007 ratio with  electron 
temperature $T_e$.The position of the Lynx arc is indicated with a solid
circle. This diagram implies T$_e\sim$17300$^{+500}_{-700}$ K  for this object.}
\end{figure}

\subsubsection{Electron density}

There are three UV line doublets we can use to estimate the electron density in the Lynx
arc: CIII]$\lambda\lambda$1907,1909, SiIII]$\lambda\lambda$1883,1892 and NIV$\lambda\lambda$1483,1487
(see \cite{keenan92}, \cite{keenan95}).
 The  results  shown in Table 1 correspond to
 T$_e\sim$10000-20000 K range, where 
the ratios are  quite  insensitive
to temperature. 
The upper limits  take the errors into account (i.e. the densities shown
correspond to the minimum possible value
of the line ratio).

The SiIII] doublet density determination has to be considered carefully.
We have noticed that the predictions by Keenan, Feibleman \& Berrington  
(1992, see Fig.2) 
are actually inconsistent with the results of our photoionization models
(\S2.3). As an example, for some of the $n_e$=1000 cm$^{-3}$
models presented in \S2.3, SiIII]1883/1892=1.8, which is
above the maximum possible value ($\sim$1.5) predicted by  those authors for
this ratio. For models with $n_e$=10 cm$^{-3}$, we find that the ratio reaches 
a value of 2 in some cases.
 We also find that the ratio varies  for a fixed density (between 
1.45 and 1.8 for $n_e$=1000 cm$^{-3}$), depending on
other parameters such as the spectral energy distribution (SED) properties.

\begin{table}
\begin{tabular}{llllll}
\hline 
Ratio &  Lynx   & $n_e$    \\ 
&            &  cm$^{-3}$  \\ \hline
CIII]$\frac{1907}{1909}$        &  1.45$\pm$0.06 & $<$3200  \\  
             &          &           \\
 SiIII]$\frac{1883}{1892}$   &  1.5$\pm$0.1 & $<$2000    \\
            &          &           \\
 NIV$\frac{1483}{1487}$   &  0.7$\pm$0.1 & 2$\times$10$^4$$^{+8000}_{-10000}$ \\ \hline
\end{tabular}
\caption{Electron densities measured for the Lynx arc using the CIII], SiIII] and NIV] UV doublets.}
\end{table}

The range of densities implied by the different doublets 
 suggests that there is a gradient in the Lynx arc. The highly ionized 
region responsible for the emission of the N$^{+3}$ lines 
has  $n_e$$\sim$[0.1-1.0]$\times$10$^5$ cm$^{-3}$.
The C$^{+3}$ lines (1549,1551 \AA ) are also emitted within this region.
There is a zone of    lower ionization level, where  the C$^{+2}$, Si$^{+2}$ 
and N$^{+2}$ lines are emitted and which 
 has $n_e$$<$3200 cm$^{-3}$. 

 Usually,        
the optical estimates of electron densities of H II regions, based     
on forbidden emission line ratios, are in the range 100 to 1500        
cm$^{-3}$.  However, radio techniques have revealed the existence in the
Galaxy of very small and dense ($n_4 >$10$^4$-10$^5$ cm$^{-3}$) objects, the so-called compact or ultracompact HII regions 
(e.g. \cite{hab79}).

Density gradients in several giant and galactic HII regions have been observed
(e.g. \cite{cop00}).
 In most of these cases, the spatial variation of density
may be interpreted as a radial gradient with the density decreasing 
from the centre to the edges.
A good example is the Orion nebula. 
Based on measurements of the ratio [O II]3729/3726, Osterbrock \&  Flather (1959)  showed the existence of a steep radial density gradient in the Orion Nebula, with the electron density decreasing
from 1.8$\times$10$^4$ cm$^{-3}$  in the centre to 2.6$\times$10$^2$ cm$^{-3}$  near the
  edge.

\subsubsection{Nebular abundances}

Photoionization models by \cite{fos03} imply
 that the nebular abundances in the Lynx arc
are $\sim$5\% solar within a factor of 2. We set here tighter constraints
by  means of using standard techniques of nebular abundance determination.

The following expressions were used to calculate the ionic abundances
of several elements (Aller 1984),  assuming the low density regime.
These expressions  will then be used 
 to calculate total element abundances.

\vspace{0.2cm}

$$\frac{O^{+2}}{H^+} = 1.12\times10^{-6}\sqrt{t} ~E^{0}_{4,2} ~\times 10^{1.25/t} ~\frac{[OIII]5007,4959}{H\beta}$$

or, alternatively

$$  \frac{O^{+2}}{H^{+}} = 7.29 \times 10^{-7} \sqrt{t} ~E^{0}_{4,2} ~\times 10^{3.75/t} ~\frac{OIII]1661,1666}{H\beta}$$

$$\frac{Ne^{+2}}{H^+} = 1.99 \times 10^{-6} \sqrt{t} ~E^{0}_{4,2} ~\times 10^{1.59/t} ~\frac{[NeIII]3869}{H\beta}$$


$$  \frac{C^{+3}}{H^{+}} = 2.04 \times 10^{-8} \sqrt{t} ~E^{0}_{4,2} ~\times 10^{4.03/t} ~\frac{CIV 1548,1550}{H\beta} $$

$$  \frac{C^{+2}}{H^{+}} = 1.11 \times 10^{-7} \sqrt{t} ~E^{0}_{4,2} ~\times 10^{3.28/t} ~\frac{CIII]1906,1909}{H\beta} $$

$$  \frac{N^{+4}}{H^{+}} = 2.17 \times 10^{-8} \sqrt{t} ~E^{0}_{4,2} ~\times 10^{5.04/t} ~\frac{NV 1239,1243}{H\beta}  $$

$$  \frac{N^{+3}}{H^{+}} = 1.06 \times 10^{-7} \sqrt{t} ~E^{0}_{4,2} ~\times 10^{4.20/t} ~\frac{NIV] 1487}{H\beta} $$

$$  \frac{N^{+2}}{H^{+}} = 2.99 \times 10^{-7} \sqrt{t} ~E^{0}_{4,2} ~\times 10^{3.57/t} ~\frac{NIII]1750_{mult}}{H\beta} $$

where  $t$=$T_e$/10$^4$ and $E^{0}_{4,2} = 1.387 ~t^{-0.983} \times 10^{-0.042/t}$

\vspace{0.2cm}

For silicon we have used the expression:

$$  \frac{Si^{+2}}{C^{+2}} = 0.188 ~t^{0.2}~e^{0.08/t} \frac{SiIII]1883,1892}
{CIII]1906,1909}$$

(from \cite{garnett95})

 Garnett et al. (1995) 
showed that  this expression
  is a  good approximations within
the temperature range 10000-20000 K and for the $n_e$ values  expected
for the Lynx arc, well below the critical densities of all the lines considered above.

The total element abundances are then calculated as follows:

$$\frac{O}{H} \sim \frac{O^{+2}}{H^+} ~~~ ~~~ ~~~ \frac{Ne}{H} \sim \frac{Ne^{+2}}{H^+}$$

In both cases, we assume that most oxygen and neon atoms are twice
ionized and therefore the  ion abundance ratio is a good representation of
the element abundance ratio. This assumption is supported by the high level of
ionization of the gas implied by the high [OIII]/H$\beta$ and
[OIII]/[OII] ratios of the Lynx arc. It is also
confirmed by our photoionization models (see \S2.2) which show that
$\sim$99\% of O and Ne are in the O$^{+2}$ and Ne$^{+2}$ forms.

For the other elements we use:

$$\frac{C}{H} \sim \frac{C^{+3}+C^{+2}}{H^+}$$

 $$\frac{N}{H} \sim \frac{N^{+4}+N^{+3}+N^{+2}}{H^+}\sim  \frac{N^{+3}+N^{+2}}{H^+}$$

(as Table 1 shows, the abundance of N$^{+4}$ in negligible compared with that of N$^{+3}$ and N$^{+2}$)

 We assume  that the low ionization species  N$^{+}$, N$^{0}$, C$^{+}$ and C$^{0}$ are not present, which is expected due to 
 the high level of ionization of the gas. It is also
confirmed by our photoionization models (see \S2.2), that also predict
negligible abundance for ionization species higher than the ones considered
above.

To calculate the abundance of Si we need to account for the presence
of different ions of this element using an ionization correction factor $ICF$:

$$\frac{Si}{C} = \frac{Si^{+2}}{C^{+2}} \times [\frac{X(Si^{+2})}{X(C^{+2})}]^{-1} 
= \frac{Si^{+2}}{C^{+2}} \times ICF_{Si/C}$$

Garnet et al. (1995) predicted the values of ICF$_{Si/C}$ for different X(O$^{+2}$) 
(see Fig. 1 in their paper). The scatter at large X(O$^{+2}$)($\sim$0.99) values implies 
that ICF$_{Si/C}$ is in the range 2.5 - 5. Our photoionization models (\S2.2) imply ICF$_{Si/C}$ in the range
3.2-3.4 (best models with no WR stars) and 2.3-2.9 (best models with WR stars). We therefore assume
   ICF$_{Si/C}$ in the range 3.4 - 2.3 (accounting for different
stellar metallicities/ages).

  The results of the
calculations are shown in Table 2. Errors were calculated using the uncertainties on the line measurements presented in Table 3\footnote{Errors on line measurements 
 provided by A. Humphrey, private communication. We have not added an error for CIV/H$\beta$, since this value is
 determined by the uncertainties on the reconstruction model of the line profile, rather than the line flux measurements (\cite{fos03})}. The errors on the abundances
relative to the Sun are less than few per cent.  C and O solar abundance values come from
photospheric abundances of Allende-Prieto, Lambert \& Asplun  (2002, 2001), while  N, Ne, Mg, Si and Fe are from 
\cite{holw01}. Uncertainties on these values were not considered in the error calculation.

\begin{table}
\begin{tabular}{llll}
\hline 
Ion &  Lynx      & Lynx          \\ 
  &  $t$=1.7 & t=$1.6$    \\ \hline
 ~O$^{+2}$/H$^{+}_{~~~~4959,5007}$    &  (5.4$\pm$0.2)$\times$10$^{-5}$  & (6.0$\pm$0.2)$\times$10$^{-5}$   \\
~O$^{+2}$/H$^{+}_{~~~~1661,1666}$    &  (6.8$\pm$0.5)$\times$10$^{-5}$   & (9.6$\pm$0.7)$\times$10$^{-5}$  \\
 ~N$^{+4}$/H$^{+}$    & $\leq$9.5$\times$10$^{-7}$  &    $\leq$1.4$\times$10$^{-6}$  \\ 
~N$^{+3}$/H$^{+}$    &   (8.2$\pm$0.2)$\times$10$^{-5}$ & (1.2$\pm$0.3)$\times$10$^{-5}$    \\ 
~N$^{+2}$/H$^{+}$    & (7.1$\pm$0.8)$\times$10$^{-6}$   &  (10$\pm$1)$\times$10$^{-6}$   \\ 
~C$^{+3}$/H$^{+}$    &  2.1$\times$10$^{-6}$  &  3.1$\times$10$^{-5}$   \\ 
~C$^{+2}$/H$^{+}$    &  (5.8$\pm$0.5)$\times$10$^{-6}$   & (7.8$\pm$0.7)$\times$10$^{-6}$   \\ 
~Ne$^{+3}$/H$^{+}$    & (1.2$\pm$0.1)$\times$10$^{-5}$     & (1.4$\pm$0.1)$\times$10$^{-5}$  \\ 
~Si$^{+2}$/H$^{+}$    &   (1.5$\pm$0.4)$\times$10$^{-6}$   &  (2.1$\pm$0.5)$\times$10$^{-6}$   \\ \hline
Element & Lynx  & Lynx \\ 
 &  $t$=1.7 & t=$1.6$ \\ \hline
~O/O$_{\odot}{_{~~~~4959,5007}}$    &  0.11  & 0.12 \\
~O/O$_{\odot}{_{~~~~1661,1666}}$    &  0.14 & 0.18  \\
~C/C$_{\odot}$    &   0.12 &  0.16 \\
~N/N$_{\odot}$   & 0.20   & 0.27 \\
~Ne/Ne$_{\odot}$   & 0.12  & 0.14  \\
~Si/Si$_{\odot}$  &  0.15-0.11(WR) & 0.21-0.14(WR) \\  \hline
\end{tabular}
\caption{Abundances  of some ions and elements in the Lynx arc for two different $t$ values 1.6 and 1.7.
The range given for Si  corresponds to the extreme values
of the ionization  correction factor $ICF$ implied by our photoionization
models (see text). The second value corresponds to models with WR stars. The two
values given for oxygen correspond to the calculations performed using the
[OIII]$\lambda\lambda$4959,5007 and the [OIII]$\lambda\lambda$1661,1666 doublets. The errors on the abundances
relative to the Sun are less than few per cent. This  only  considers  the errors on the ionic abundances and do not account for 
uncertainties on the Solar values.}
\end{table}

Calculations were done for two temperature values $T_e$=16 000 and 17 000 K ($t$=1.6 and 1.7) 
to account for the most likely $T_e$ range implied by the photoionization models (\S2.2) and the  
 [OIII] lines (\S2.1.1).  It is important to note that the electron
 temperature is different in high and low ionization
zones of HII regions. To perform a more accurate
abundance determination, this temperature variation should be taken
into account.  Most of the gas is ionized in the Lynx arc ($\sim$99 \% of oxygen in  the O$^{+2}$ form) and there
is not a low ionization zone  (i.e. a region where species such as
[OI], [OII], [SII], etc exist). We expect therefore that
the electron temperatures 
  measured using the [OIII] lines can be safely 
assumed to be representative of a large fraction of the nebula.
Somewhat higher temperatures
might be present in the most highly ionized region. In such case, the 
abundances presented in Table 2, are  upper limits
\footnote{The small discrepancy between the O/H values calculated with the
[OIII]$\lambda\lambda$5007,4959 and OIII]$\lambda\lambda$1661,1666
values (see Table 2) could  be due the fact that  the excitation of the UV doublet
requires  somewhat higher electron temperatures than
the optical doublet.  }.

We conclude that the nebular metallicity  in the Lynx arc (defined as O/H)
is in the range $\sim$10 - $<$20\% Z$_{\odot}$. This is somewhat higher than the value derived
by \cite{fos03} of $\sim$5\%  Z$_{\odot}$.

In addition,  we conclude that nitrogen is  overabundant relative to oxygen
 by a factor of $\sim$2.0-3.0  compared with the solar abundance.

Departures from the constant density case
may produce variations in the derived ICF values and therefore, the
element abundances (Luridiana \& Cervi\~no 2003, Luridiana et al. 2004, in prep.). This does not
affect oxygen seriously. However,  the derived abundance of nitrogen
 can be off by up to a factor
of 1.75 for their models with the steepest power law density gradient (index $>$1.5).
It is  therefore possible, that  N is not overabundant and the large
derived N/O is a consequence of
assuming
constant density.

\cite{fos03} concluded that silicon has to be well overabundant 
relative to other elements,
to be able to explain the strong SiIII] lines detected in the Lynx arc. 
The reason for this discrepancy will be discussed in \S2.2.

Photoionization models in \S2.2 will set further constrains on the nebular abundances
of all elements.

\subsection{Age and metallicity of the ionizing stellar population}

We have characterized the age and metallicity of the ionizing stellar
population in the Lynx arc by means of producing photoionization models
that best reproduce the emission line properties.
Further constraints on the nebular physical properties and chemical abundances
will also be set.

The most outstanding spectroscopic properties of the Lynx arc  that
 models should reproduce are (see \cite{fos03}):

\begin{itemize}

\item The large strength of the UV collisionally excited lines 
(NIV$\lambda$1486, CIV$\lambda$1550, OIII]$\lambda$1663, NIII]$\lambda$1750, etc) relative to recombination lines such as H$\beta$

\item The large [OIII]/H$\beta$(=7.5) and the weakness of [OII]$\lambda$3727 relative to other lines
([OIII]/[OII] $\geq$30, [OII]/H$\beta$$\leq$0.25  and [NeIII]$\lambda$3869/[OII]$\geq$2.8)). I.e., the high
ionization level of the gas.

\item The strength of the He$^{+2}$ emission (HeII$\lambda$1640/H$\beta$=0.11)

\item The extremely weak UV and optical continuum

\end{itemize}

All the SEDs used in this paper were  computed by Cervi\~no, Mas-Hesse \& Kunth (2004, hereafter CMK04)
available at {\tt http://www.laeff.esa.es/users/mcs/SED}.
 The isocrones were computed following the prescriptions quoted in \cite{cer01} and the models assume a power 
law Initial Mass Function (IMF) with a Salpeter slope, $\alpha$=2.35 (\cite{Salpeter55}) in the mass range 2--120 M$_\odot$. We have adopted the evolutionary
tracks with standard mass loss rates by \cite{Schetal92,Schetal93a,Schetal93b,Charetal93} and the following atmosphere models: \cite{CoStar} {\sc CoStar} for main-sequence hot stars more massive than 20 M$_{\odot}$, \cite{Schmetal92} for WR stars and
\cite{kur} for the remaining stars.
 An instantaneous star formation law was assumed.
 We have also considered the influence of the X-ray emission of the starburst
 in the  SED by varying the efficiency of conversion of kinetic energy into X-rays. However, 
we have found that this effect does not produce significant differences in the
predicted emission line spectrum at the young ages considered here. Therefore, we have ignored it
and  we present only the results for SED models without X-ray emission.

 We used   the numerical code Cloudy (\cite{fer98}) to predict the line
ratios in different circumstances.
 
Given the complete lack of information on the geometry of the gas and the relative distribution between stars and gas in the arc, we have assumed the
 standard  spherical geometry often used in
photoionization 
models of other extragalactic
HII regions and HII galaxies. The distance between the ionizing source and the gas $r_0$  was assumed to be 10 pc  and we used constant density $n$=1000 cm$^{-3}$. The filling
factor $ff$ was varied until finding the model that best reproduces the Lynx arc
line ratios.
 Conclusions on the size and geometry of the ionized
region cannot be extracted from our modeling since different values of $ff$,  $n$ and
$r_0$ might produce the same output spectrum. In addition, the assumption of spherical geometry is likely
to be too simple, since the ionizing stars might be distributed in several clusters.

We adopt log $Q(H)$ = 55.20, as calculated by \cite{fos03} from the H$\beta$ flux
 and corrected
for magnification. $Q(H)$ is the ionizing luminosity in erg s$^{-1}$. Since the authors assumed 
covering factor $CC$=1 in these calculations,
we have also adopted this value.

We show in Tables 3 to 6 the results of the models that {\it best} 
reproduce the Lynx spectrum,  together with the Lynx measurements. Although He$^{+2}$ is a problem
for some of these models, they are also shown  because of the very good agreement with the rest of
the line ratios. The He$^{+2}$ problem will  be discussed in more detail in \S3.1. 
 We find that the nebular metallicity Z$_{neb}$ that produces
the best fit to the Lynx spectrum is $\sim$10\% Z$_{\odot}$.  This is in very good agreement
with the results obtained in \S2.1.3. 

 Z$_{neb}$=5\% Z$_{\odot}$ models
are rejected.  The models that can 
produce [OIII]5007/H$\beta\ga$7, require very high filling factor $ff\ga$0.32
(i.e. high ionization level).
The predicted electron temperature is $\sim$20000 K, much higher than the
value implied by the [OIII] lines (\S2.1.1). In addition, due to the high $T_e$ and ionization
level of the nebula,
 the high ionization UV lines such as CIV$\lambda$1550 and [OIII]$\lambda$1665
 become too strong, with CIV/H$\beta>7$.

The opposite effect is found for models
with  Z$_{neb}\ge$15\% Z$_{\odot}$,
which can also be  rejected with confidence. Due to the higher
metallicity, lower filling factors (i.e. ionization level) 
are required to produce the same [OIII]$\lambda$5007/H$\beta$.  For those models 
with 
[OIII]5007/H$\beta\ga$7-8, the predicted $T_e$ is $\sim$14 000K, too low compared with
the expected value (\S2.1.1).
In addition, because of the low ionization level and the low electron temperature, the high ionization UV collisionally excited lines become too weak, with CIV/H$\beta\la$1.

Therefore, all the models presented here assume Z$_{neb}$=10\% Z$_{\odot}$ for
all elements (except nitrogen, see below).  The most discrepant line ratios will always
be shown in bold characters.

\vspace{0.2cm}

\centerline{Z$_{stars}$ =  5\% Z$_{\odot}$ models}

\vspace{0.2cm}

The results of these models are shown in Table 3. 
The 2 Myr($a$) model assumes that all elements have an abundance of
10\% the solar value.
Notice that the NIII] and NIV lines are   severely underestimated. This is the case
for all models considered below, as long as N abundance is not enhanced. 
The discrepancy disappears if we assume that this element is overabundant by a factor of $\sim$3.
This result was also suggested by   the nebular abundance determination in \S2.1.3.
Hereafter, this overabundance  will be assumed in all models.

Models  with ages $\leq$4 Myr reproduce very well the Lynx line properties, except the He$^{+2}$ emission. 
Ages of 5 Myr and specially older present strong discrepancies with the measured line ratios,
 due to the drastic reduction of the   ionizing photons as stars get older.
We only show 5 and a 6 Myr models in Table 3 for illustration.

    The  He$^{+2}$ lines  are  always
predicted too faint. The reason is that according to the stellar evolution models
no WR stars are formed in such low metallicity environment (but see \S3.1).

\vspace{0.2cm}

\centerline{Z$_{stars}$ =  20\% Z$_{\odot}$ models}

\vspace{0.2cm}

Models with ages $\la$3 Myr  
are in very good agreement with the  Lynx spectrum (Table 4), except for the He$^{+2}$ lines which are predicted
too faint. At $\sim$3.4 Myr Wolf Rayet stars  make an important contribution to
the the hard ionizing radiation and the He$^{+2}$ problem is solved. 
 NV should be detected according to this model.
However, given the uncertainties on the estimation of the upper limit for this line, the predicted ratio
 is likely to be well within the errors (\cite{fos03}) of the measured value.
Ages  of 4 Myr and older  produce strong discrepancies with the measured spectrum.
 We  show  4 and  5 Myr models in Table 4 for illustration.

\vspace{0.2cm}

\centerline{Z$_{stars}$ =  40\% Z$_{\odot}$ models}

\vspace{0.2cm}

Models  with 
ages $<$3 Myr  are in good agreement with the Lynx emission line spectrum, except for the  He$^{+2}$ lines which are predicted 
too weak (Table 5).
Between 3.0 and 4 Myr, WR stars appear and the He$^{+2}$ problem is solved.
Ages of 4 Myr and older produce strong discrepancies with the measured spectrum.  We  show 4     and    5 Myr models
 in Table 5 for illustration.

\vspace{0.2cm}

\centerline{Z$_{stars}$ =  Z$_{\odot}$ models}

\vspace{0.2cm}

Models with ages  $\la$5 Myr can explain the Lynx emission line spectrum remarkably well, except the He$^{+2}$  lines, which
are predicted too weak.
 The appearance of WR stars at 2.8 Myr solves 
the  He$^{+2}$  problem and produces
a  good fit to the observed spectrum. Ages of 6 Myr and older produce strong discrepancies with the measured spectrum. 
We only show a  6 Myr model  in Table 6 for illustration.

\begin{table*}
\begin{tabular}{llllllllll} \hline
 ~	    Z$_{stars}$ /Z$_{\odot}$     &  Lynx   &   5\%    & 5\%         & 5\%     &  5\%   &  5\%   &   5\%    &   5\%     \\
~	    Age$_{stars}$           &     & 1 Myr      & 2 Myr$_a$        & 2 Myr$_b$      &    3 Myr & 4 Myr & 5 Myr   &   6 Myr      \\
~  Z$_{neb}$/Z$_{\odot}$              &    &   10\%            &  10\%        &     10\%  &        10\% &  10\%   &  10\%   &  10\%      \\
~  N/N$_{\odot}$         &          &  30\%  &  10\%       &     30\%       &    30\%   &      30\%   &      30\% &      30\%   \\ 
 ~ log($ff$)            &                  & -1.7 & -1.5   & -1.5 & -1.3  & -0.3   & 0.0   & 0.0      \\ \hline
 ~ NV$\lambda$1240           &   $\leq$0.09     &  0.01 &0.01    &  0.01      & 0.03   &       0.01 & 0.13 & {\bf 0.27} \\
~  SIV$\lambda$$\lambda$1394  &   $\leq$0.09    &  0.13 & 0.12   & 0.12    & 0.10 &  0.12  &  0.13 & 0.09 \\
~   SIV$\lambda$$\lambda$1402   &   $\leq$0.09  &  0.06    &  0.06 &0.06       & 0.06   & 0.06 & 0.10 & 0.07   \\
 ~ NIV]$\lambda$1486         &   0.42$\pm$0.05           & 0.52 & {\bf 0.18}     &  0.52  &  0.47    & 0.52   &  0.28 & {\bf 0.06}      \\
 ~ CIV$\lambda$1549          &   3.65        & 3.63   &  3.90    & 3.69    & 3.33      & 3.63 &  {\bf 1.97}  & {\bf 0.60}    \\
 ~ HeII$\lambda$1640         &   0.11$\pm$0.03     &  {\bf 0.008} &  {\bf 0.008}  &  {\bf 0.008}      &   {\bf 0.004}    &  {\bf 0.005}    &{\bf 0.012}     & {\bf 0.023}     \\
 ~ OIII]$\lambda$1665        &   0.56$\pm$0.04           & 0.52 &  0.54       & 0.52       &  0.46     &  0.52   &  0.37   & {\bf  0.17} \\
 ~  NIII]$\lambda$1750       &   0.18$\pm$0.02      & 0.17 &  {\bf 0.05}       & 0.17      &  0.13     &  0.17       &  0.17  &  0.13 \\
 ~ SiIII]$\lambda$1883       &   0.09$\pm$0.02     & 0.12      &  0.11         &  0.11    &  0.11    &   0.12    & {\bf  0.03} & 0.11  \\
 ~ SiIII]$\lambda$1892       &   0.06$\pm$0.02     & 0.07      &  0.05         & 0.07      & 0.07      & 0.07    &  {\bf  0.02} & 0.08  \\
 ~ CIII]$\lambda$1909        &   0.59$\pm$0.06           & {\bf 0.94} &  {\bf 0.88}     & {\bf 0.91}      &  0.74     & {\bf 0.99}     & {\bf 0.84}  & {\bf 0.84}   \\
 ~ [OII]$\lambda$3727       &    $\leq$0.25     &  0.08 & 0.08         &  0.08     &  0.06      & 0.08    & 0.03   & {\bf 0.43}  \\
 ~ [NeIII]$\lambda$3869      &   0.69$\pm$0.06           &  0.62 & 0.63         &   0.62    & 0.59      & 0.62    & 0.58 &  {\bf 0.37}  \\
 ~ [NeIII]$\lambda$3968      &   $\leq$0.22     &  0.19 & 0.19         &   0.19   &  0.18    &   0.19  &  0.17 &   0.11   \\
 ~  HeII $\lambda$4686   &  0.015-0.025    &  {\bf 0.0011} &  {\bf 0.0011}  & {\bf 0.0011}      &  {\bf 0.0006}       &  {\bf 0.0007}  &  {\bf 0.0016}   & {\bf 0.0024} \\
 ~ [OIII]$\lambda$5007       &  7.50$\pm$0.3   & 8.25          &   8.36      &  8.28         &  7.97    & 8.25 & 7.8  & {\bf 5.43}  \\ 
 ~ H$\beta$    &      1.00    & 1.00  & 1.00  &  1.00       &  1.00   &  1.00   &  1.00  &  1.00          \\  \hline
 ~ $t=\frac{T_e}{10 000}$                  &  1.73$^{+0.50}_{-0.70}$        & 1.64  &   1.64   &  1.65   &  1.59 &    1.64  & {\bf 1.48}  & {\bf 1.31}    \\    \hline
\end{tabular}
\caption{Z$_{stars}$= 0.001 = 5\% Z$_{\odot}$.  Line ratios are always given relative to H$\beta$. Discrepant values are shown in
bold. The NV, NIV], CIV, CIII], OIII]1665 and [OII] values  include the fluxes of the two doublet lines. 
 The agreement between the models with ages of 4 Myr or less
 and the observations is remarkably good, except for the  He$^{+2}$ lines, which are always predicted too faint. Considering 
N/O $\sim$ (N/O)$_{\odot}$, the models underestimate the strength of the nitrogen emission lines (see model 2 Myr$_a$). The problem disappears when
nitrogen is assumed to be overabundant relative to oxygen by a factor of $\sim$3. Models of  5 Myr and older
are  inconsistent with the Lynx arc
spectrum. We only show a 5 Myr and a 6 Myr model for illustration.}
\end{table*}

\begin{table*}
\begin{tabular}{lllllllll} \hline
~	    Z$_{stars}$ /Z$_{\odot}$     &   Lynx       & 20\%         & 20\%     &  20\%    &  20\%   &  20\%    &  20\%       \\
~	    Age$_{stars}$           &           & 1 Myr      & 2 Myr      &    3 Myr & 3.4 Myr &  4 Myr  &  5 Myr       \\
~  Z$_{neb}$/Z$_{\odot}$              &          &  10\%        &     10\%       &  10\%  &  10\% &  10\%   &  10\%     \\
~  N/N$_{\odot}$         &          &  10\%        &     30\%       &    30\%    &    30\% &    30\%  &    30\%      \\ 
 ~ log($ff$)                        &                  &  -1.7  & -1.5 &    -0.5 & -1.5 &  0.0  &  0.0      \\ \hline
 ~ NV$\lambda$1240           &   $\leq$0.09     &  0.012   &  0.02     &  0.05        &   {\bf 0.16} &  0.12  & {\bf 0.18}    \\
~  SIV$\lambda$$\lambda$1394  &   $\leq$0.09    &  0.14    & 0.12         & 0.09 &  0.11 & 0.08 & 0.07        \\
~   SIV$\lambda$$\lambda$1402   &   $\leq$0.09  &  0.07    & 0.06         & 0.05  & 0.06  &     0.05 &  0.05     \\
 ~ NIV]$\lambda$1486         &   0.42           &  0.54    &  0.47  &  0.46    &  0.53  &    0.44  & {\bf 0.05}     \\
 ~ CIV$\lambda$1549          &   3.65           &  3.79    & 3.25    & 3.25       & 3.61  & 3.16   &  {\bf 0.55}    \\
 ~  HeII$\lambda$1640         &   0.11     & {\bf 0.007}   &  {\bf 0.006}     &  {\bf 0.007}    & 0.16  & {\bf 0.009} &  {\bf 0.016}              \\
 ~ OIII]$\lambda$1665        &   0.56           &  0.54       & 0.49       &   0.43    & 0.47  & 0.39    & {\bf  0.13}     \\
 ~  NIII]$\lambda$1750       &   0.18      &  0.17         & 0.17      &  {\bf  0.11}      & 0.13   &  {\bf  0.09 }  &  {\bf 0.10}       \\
 ~ SiIII]$\lambda$1883       &   0.09           &  0.11         & 0.12     &  0.04     &  0.11  &  {\bf  0.03}  & {\bf  0.13}    \\
 ~ SiIII]$\lambda$1892       &   0.06           &  0.07         & 0.07      &  0.03     &  0.06  &   {\bf  0.02}&  {\bf  0.08}    \\
 ~ CIII]$\lambda$1909        &   0.59           &  {\bf 0.95}    &    {\bf 0.94}      &  0.63        & 0.72   & 0.51 & 0.80       \\
 ~ [OII]$\lambda$3727       &    $\leq$0.25     &  0.08         & 0.09      &  0.04     & 0.10    &  0.03  & {\bf 0.79}  \\
 ~ [NeIII]$\lambda$3869      &   0.69           &   0.63        &  0.60     & 0.59      & 0.58  &  0.59  & {\bf 0.34}  \\
 ~ [NeIII]$\lambda$3968      &   $\leq$0.22     &  0.19         & 0.19     &  0.18     & 0.17  & 0.18    & {\bf 0.10}   \\
 ~  HeII $\lambda$4686  &  0.015-0.025    &  {\bf 0.0011}  &  {\bf 0.0011}     &  {\bf 0.001}    &  0.02      & {\bf 0.0012} & {\bf 0.0017}  \\
~ [OIII]$\lambda$5007       &  7.50             & 8.24             & 7.98     &   7.98   &  7.60  &  8.01  &  {\bf 4.92}     \\ 
 ~ H$\beta$    &      1.00    & 1.00  & 1.00  &  1.00       &  1.00    & 1.00  & 1.00     \\  \hline
 ~ $t$                  &    1.73$^{+0.50}_{-0.70}$ &  1.67     &   1.62  &  {\bf 1.56}  & 1.62  &   {\bf 1.50}  & {\bf 1.28}         \\  
 					    &          &   &    &          &    {\bf  WR }       &     \\  \hline
\end{tabular}
\caption{Z$_{stars}$= 0.004 = 20\% Z$_{\odot}$. All models assume N/O $\sim$ 3 $\times$ (N/O$_{\odot}$). In spite of the He$^{+2}$  problem (see text) the agreement between the models (age 3  Myr or less)
 and the observations is remarkably good. The He$^{+2}$ problem is solved at $\sim$3.4 Myr, when
Wolf Rayet (WR) stars appear and ionize  He$^{+}$. Models of 4 Myr and older are  inconsistent with the Lynx arc
spectrum. We only show a 5 Myr old model for illustration.}
\end{table*}

\begin{table*}
\begin{tabular}{llllllll} \hline
 ~	    Z$_{stars}$ /Z$_{\odot}$     &    Lynx      & 40\%         & 40\%     &  40\%   &  40\%   &  40\%  &  40\%      \\
~	    Age$_{stars}$           &          & 1 Myr        & 2 Myr      &    3 Myr & 3.3 Myr &   4 Myr &  5 Myr       \\
~  Z$_{neb}$/Z$_{\odot}$              &          &  10\%        &     10\%       &  10\%   &  10\%  &  10\%   &  10\%    \\
~  N/N$_{\odot}$         &          &  10\%        &     30\%       &    30\%   &    30\%  &    30\%   &    30\%   \\ 
 ~  log($ff$)                       &                  &  -1.7  & -1.3   &  -1.3 & -1.0 &   0.0  &   0.0   \\ \hline
 ~ NV$\lambda$1240           &   $\leq$0.09     & 0.01    & 0.022      &   0.12       &  {\bf 0.16} & {\bf 0.20} & {\bf 0.24}   \\
~  SIV$\lambda$$\lambda$1394  &   $\leq$0.09    & 0.12     &   0.11       & 0.11 &  0.09  & 0.06  &      0.08   \\
~   SIV$\lambda$$\lambda$1402   &   $\leq$0.09  & 0.06     &   0.06       & 0.06  & 0.05  &  0.04 &      0.07  \\
 ~ NIV]$\lambda$1486         &   0.42           &  0.45    & 0.48   & 0.53     & 0.52   &  0.51  & {\bf 0.03} \\
 ~ CIV$\lambda$1549          &   3.65           &  3.13    & 3.34    &  3.69     & 3.69  & 3.77  & {\bf 0.50}    \\
 ~  HeII$\lambda$1640        &   0.11     &{\bf  0.007}     &   {\bf  0.006}     &  0.10    & 0.14  & {\bf 0.18}  & {\bf  0.02}       \\
 ~ OIII]$\lambda$1665        &   0.56           &  0.48       & 0.48       &   0.48    & 0.44    &  0.40  & {\bf  0.07}  \\
 ~  NIII]$\lambda$1750      &   0.18      &  0.18         &  0.15     &  0.14     &   {\bf 0.09}      &  {\bf 0.04}& {\bf  0.06}  \\
 ~ SiIII]$\lambda$1883       &   0.09           &  0.13         &  0.10    &  0.09     &  0.06     & {\bf  0.018} & {\bf 0.19  }  \\
 ~ SiIII]$\lambda$1892       &   0.06           &  0.08         &  0.06     & 0.06      & 0.04    & {\bf 0.013} & {\bf 0.12 }  \\
 ~ CIII]$\lambda$1909        &   0.59           &   {\bf 0.98}        & {\bf 0.84}      & 0.70      &  0.50    &   {\bf 0.21}  &      0.77     \\
 ~ [OII]$\lambda$3727       &    $\leq$0.25     &   0.09        &  0.08     & 0.08      &  0.06   & 0.03 & 2.17 \\
 ~ [NeIII]$\lambda$3869      &   0.69           &  0.60         &  0.60     & 0.59      & 0.57  &    0.57    & {\bf  0.20} \\
 ~ [NeIII]$\lambda$3968      &   $\leq$0.22     &  0.18         &  0.18    &  0.18     & 0.17 &   0.17  & 0.06 \\
 ~  HeII $\lambda$4686   &  0.015-0.025    &  {\bf 0.001}   &   {\bf 0.0008}    & 0.012     & 0.017    & 0.022  & {\bf 0.002} \\
 ~ H$\beta$                  &  1.00           &   1.00        & 1.00  & 1.00  &  & 1.00  & 1.00    \\
~ [OIII]$\lambda$5007       &  7.50             & 7.95    &  7.97    &  7.85    &  7.59    &  7.8 &  {\bf 3.01 }   \\ 
 ~~ H$\beta$    &      1.00    & 1.00  & 1.00  &  1.00       &  1.00   & 1.00   & 1.00    \\ \hline
 ~ $t$                  &    1.73$^{+0.50}_{-0.70}$       &   1.62    &  1.61   & 1.62  &     1.59   & {\bf 1.52} & {\bf 1.20}   \\ 
 					    &          &   &    & {\bf  WR }          &    {\bf  WR }       &  {\bf  WR }   \\  \hline
\end{tabular}
\caption{Z$_{stars}$= 0.008 = 40\% Z$_{\odot}$. All models assume N/O $\sim$ 3 $\times$ (N/O$_{\odot}$).  In spite of the He$^{+2}$   problem (see text) the agreement between the models
and the observations is remarkably good for an age of 3.3 Myr or less. The best models are those with age $\sim$3 Myr, when
WR stars appear and ionize He$^{+}$. Models of 4 Myr or more  produce strong discrepancies with the observations.}
\end{table*}

\begin{table*}
\begin{tabular}{llllllll} \hline
 ~	    Z$_{stars}$ /Z$_{\odot}$     &  Lynx        & 100\%            & 100\%     &  100\%     &  100\%   &  100\%     &  100\%        \\
 ~	    Age$_{stars}$           &         & 1 Myr  & 2 Myr           & 2.8 Myr     & 4 Myr    &   5 Myr &   6 Myr   \\
~  Z$_{neb}$/Z$_{\odot}$              &          &  10\%        &     10\%       &  10\%    &  10\%  &  10\% &  10\%    \\
~  N/N$_{\odot}$         &          &  10\%        &     30\%       &    30\%  &    30\%   &    30\%   &    30\%     \\ 
 ~    log($ff$)                    &                  & -1.3   &-0.5  & -1.3  &   -1.3    & -0.7 & 0.0  \\ \hline
 ~ NV$\lambda$1240           &   $\leq$0.09     & 0.02    &  0.04 &    0.15    &   {\bf 0.32} & 0.16    & {\bf 0.58}  \\
~  SIV$\lambda$$\lambda$1394  &   $\leq$0.09    &  0.12   &  0.09        & 0.10  &  0.10  &  0.10  & {\bf 0.16} \\
~   SIV$\lambda$$\lambda$1402   &   $\leq$0.09  &  0.06    &  0.05        &   0.06 & 0.06  & 0.06 & {\bf 0.14} \\
 ~ NIV]$\lambda$1486         &   0.42           &  0.47    &  0.49  &   0.51    &  0.52  &  0.52  & {\bf 0.046} \\
 ~ CIV$\lambda$1549          &   3.65           &  3.26    & 3.48  &  3.57    &   3.73    &  3.9    &   {\bf 0.79}     \\
 ~  HeII$\lambda$1640         &   0.11     & {\bf 0.012}   & {\bf 0.009}      &  0.13      & {\bf 0.26}   &   {\bf 0.038}    &    0.071    \\
 ~ OIII]$\lambda$1665        &   0.56           &   0.48      &  0.45      &  0.45     &  0.42   &  0.44 & {\bf 0.06}   \\
 ~  NIII]$\lambda$1750       &   0.18      &  0.16         &   0.11    &  0.12     &   {\bf 0.08}       & {\bf 0.08} & {\bf 0.04}   \\
 ~ SiIII]$\lambda$1883       &   0.09           &  0.10        & 0.09     &    0.09    &    0.08   &  0.05  &  {\bf 0.17}    \\
 ~ SiIII]$\lambda$1892       &   0.06           &  0.06         & 0.05      &    0.06    &   0.05  & 0.03 &  {\bf 0.14} \\
 ~ CIII]$\lambda$1909        &   0.59           & {\bf 0.90}    & 0.63  &  0.62    &  0.47     &  0.44   &  0.49    \\
 ~ [OII]$\lambda$3727       &    $\leq$0.25     &  0.08         &  0.04     &  0.09     &  0.09   & 0.05 & {\bf 2.81} \\
 ~ [NeIII]$\lambda$3869      &   0.69           &  0.60         &  0.61     &  0.56      & 0.53 & 0.57  & {\bf 0.12}   \\
 ~ [NeIII]$\lambda$3968      &   $\leq$0.22     &  0.18         &  0.18    &   0.17      &  0.16  & 0.17 &   {\bf 0.03} \\
 ~  HeII $\lambda$4686   &  0.015-0.025    & {\bf 0.0016}  &  {\bf 0.0012}     &  0.017        & {\bf 0.033} & {\bf 0.0046} & {\bf 0.0063 }   \\
~ [OIII]$\lambda$5007       &  7.50             &   8.00        & 8.2   & 7.49    &  7.0    &   7.5   & {\bf 1.53} \\ 
~~ H$\beta$                  &  1.00           &   1.00        & 1.00  & 1.00  &  1.00 &  1.00 &  1.00  \\ \hline
 ~ $t$                  &    1.73$^{+0.50}_{-0.30}$       &   1.61    &  1.57   &   1.60  &  1.60 & 1.59   &  {\bf 1.25}        \\  
 					    &          &    &    & {\bf  WR }   &      {\bf  WR }   &         &    \\  \hline
\end{tabular}
\caption{Z$_{stars}$= 0.020 =  Z$_{\odot}$.  All models assume N/O $\sim$ 3 $\times$ (N/O$_{\odot}$).  In spite of the  He$^{+2}$ problem (see text) the agreement between  the models
and the observations is remarkably good for ages of 5 Myr or less. The best model 
corresponds to 2.8 Myr of age, when Wolf Rayet are responsible for the 
ionization of He$^+$. Models of 6 Myr or more  produce strong discrepancies with the observations. }
\end{table*}

\vspace{0.2cm}
To summarize, the best  photoionization models imply:

\begin{itemize}

\item The nebular abundances in the Lynx arc are $\sim$10$\pm$3\% Z$_{\odot}$.
This is in very good agreement with the measured abundances in \S 2.1.3 and
with \cite{fos03}  within the errors.
Lower ($<$7\% Z$_{\odot}$)  and higher ($\geq$15\% Z$_{\odot}$) metallicities
produce strong discrepancies with the measured line ratios. 

\item Nitrogen is overabundant. The models suggest N/O $\sim$ 3 $\times$ (N/O)$_{\odot}$.
This is in good agreement with the results obtained in \S2.1.3.

\item  Our models do not
imply overabundance  of Si relative to O, contrary to the results obtained
by \cite{fos03}. 
The reason to claim Si overabundance by this authors is that their model 
predicts SiIII]/H$\beta$ ratios
a factor of $\sim$100 below the measured value. Si$^{+2}$  is the species
found in the Lynx arc with the lowest ionization potential (IP). 
All other 
lines detected are emitted by ions with higher IP
(N$^{+2}$ and C$^{+2}$  are  next, with  47.4 eV and 47.9 eV respectively). 
A hot 
black body (80 000 K in \cite{fos03} models) produces 
 a large supply of ionizing
photons that efficiently remove Si$^{+2}$ while still keeping a considerable
 amount
of other species such as N$^{+2}$ and C$^{+2}$.
The same can be said about \cite{bin03} models. The ionization
of the gas in our models is dominated by much colder stars that allow the
survival of  Si$^{+2}$ in the nebula.

\item The electron temperature is $T_e\sim$16200$\pm$500 K, in good agreement
within the errors 
with the value determined from the [OIII]1663/5007 ratio (see \S2.1.1).

\item Ages $\la$5 Myr produce good fits to the Lynx spectrum, except for the He$^{+2}$  lines,
which are predicted too faint in most cases. Only when
 Wolf Rayet stars appear, the He$^{+2}$ lines are properly fit. {\it If}~ Wolf Rayet
stars are responsible for the ionization of He$^{+}$ (but see \S3.1), this implies:

\begin{itemize}

\item  Z$_{stars}$ must be $>$5\% Z$_{\odot}$, since no Wolf 
Rayet stars are formed at lower metallicities, according to the models (but see \S3.1).

\item  the Lynx arc is undergoing a WR phase and the age of the burst 
is in the range 
2.8 - 3.4 Myr, depending on the stellar metallicity. 
\end{itemize}

\end{itemize}

\section{Discussion}

\subsection{The  He$^{+2}$ problem}

All models in Tables 3 to 6 with ages  of  $\sim$5 Myr or younger (depending on the stellar metallicity)
reproduce remarkably well most line ratios of
the Lynx spectrum. Some of those models, however (those with no contribution from WR stars)
  cannot explain the   He$^{+2}$ nebular emission. 

HeII$\lambda$4686 nebular
 emission has been detected in the optical spectrum of numerous 
extragalactic HII regions and star
forming galaxies (e.g. \cite{garnett91}). If the excitation of this line is due
to photoionization by hot stars, these must have $T_{eff}\ga$55 000 K. Based on 
photoionization modeling, \cite{fos03} 
 concluded that  
$T_{eff}\sim$80 000 K in the Lynx arc, in the range of values typical of Wolf Rayet
stars.

We have shown  that Wolf Rayet stars are good candidates to explain the He$^{+2}$  emission
in the Lynx arc (they could also be responsible for  the outflowing wind discovered  
by \cite{fos03}). The non detection of the typical Wolf Rayet bumps found in  Wolf 
Rayet galaxies (e.g. \cite{vac92}) is simply due to the fact that the stellar population is not detected
at all (see \S3.3). 

We have used two of the SEDs that best reproduce the Lynx spectrum (Z$_{stars}$=20\% Z$_{\odot}$ and 3.4 Myr;
  Z$_{stars}$=40\% Z$_{\odot}$ and 3.0 Myr, see \S2.2) to predict the number of WR stars in the Lynx arc.
   CMK04 models predict 2.9$\times$10$^{-4}$ 
and  2.5$\times$10$^{-4}$ Wolf Rayet stars
per solar mass for these two   SEDs. 
Using the total stellar burst masses  estimated from the H$\beta$
luminosity (see \S3.3) ,  we obtain  2.7$\times$10$^{4}$  or
2.1$\times$10$^{4}$ WR stars for the 20\% Z$_{\odot}$ and 40\% Z$_{\odot}$ SEDs respectively.
 This is in the range of values found  for low
redshift WR galaxies, where 
the presence of 100 to 100 000 WR stars has been inferred
(\cite{vac92}).

WR stars have been proposed to explain the
He$^{+2}$ emission in stellar ionized nebulae. However, this 
is an unsolved problem in  nearby star forming objects. For this reason,
 we do not reject the possibility that
 another unknown ionization source is present. 

As an example, evidence for the existence of Wolf Rayet stars in the metal
poor galaxy I Zw 18 (Z=1/50 Z$_{\odot}$) has been reported
by several authors (\cite{izo97}, \cite{leg97}). \cite{sta99} showed that the
He$^{+2}$ nebular intensity can be reproduced using a radiation field consistent with
the observed Wolf-Rayet spectral features in this object (see also \cite{demello98}). However, it is not
clear whether WR stars can 
explain the spatial distribution of the He$^{+2}$ emission (\cite{vil98}).

Izotov et al. (2004) have recently discovered the high ionization emission line [NeV]$\lambda$3426 in the 
spectrum of the blue compact low metallicity galaxy TOL 1214-277.  Fricke et al. (2001)
 detected the [FeV]$\lambda$4227 line. 
Izotov et al. (2004) conclude that the stellar radiation is too soft to explain the existence of
these high ionization species and propose that fast shocks or high mass X-ray binary systems
could be responsible.
 The possibility that ultra-luminous X-ray sources (ULX) are responsible for 
the He$^{+2}$ emission in the dwarf irregular galaxy Holmberg II has been proposed by 
 \cite{paku02} and \cite{kaa04}. ULXs may be be accreting normal mass ($<$20 M$_{\odot}$)
black holes or  neutron stars. An alternative possibility is that ULXs are high-mass X-ray binaires with 
super-Eddington mass transfer rates (\cite{king01}).

Peimbert, Sarmiento \& Fierro (1991) showed that the emission line spectra  of giant HII regions can be
altered by the presence of shock  
waves 
produced by stellar winds or  supernova events. This is also the case in star forming
galaxies depending on the evolutionary phase of the stellar cluster
(\cite{vie00}). The outflowing wind in the Lynx arc 
discovered by \cite{fos03}  provides evidence for the possible presence of shocks
in the Lynx arc.  Such kind of effects are not included in our
synthesis models (where the supernova X-ray emission is the time-average 
contribution of the events).

A final possibility is the presence of super-soft  X-ray sources (\cite{rap94}). 
One of the physical models  proposed to explain  the nature of these sources involves
mass transfer from a main-sequence or subgiant donor star to the surface of a white dwarf in
a binary system (\cite{van92}). If this is the case in the Lynx arc, an additional stellar population 
with an age older than 50 Myrs must be present. 

Therefore, we do not discard the possibility that alternative mechanisms to WR stars are responsible for the ionization of He$^{+}$  in the Lynx arc.

\subsection{The nitrogen overabundance}

We have found that  the strength of the NIII] and NIV]  lines cannot be explained
assuming solar abundances.   As explained above, a possible
explanation for this discrepancy is the assumption of constant density, rather than taking into account
the presence of a gradient (see \S2.1.3) (Luridiana et al. 2004, in preparation).

An alternative explanation is that N is overabundant by a factor  $\sim$2-3 relative to the solar
abundance (see \S2.1.3 and \S2.2). Significant
nitrogen excesses, although rare,
have been found in some star forming objects.  Particularly interesting are the five extremely metal poor
BCGs with large nitrogen excesses (see \cite{pus04} for a more detailed
 discussion on  this issue). The authors propose that the nitrogen excesses
could be a consequence of merger events and a short powerful starburst phase
when many WR stars contribute to a fast enrichment of the ISM.

The 
morphology (\cite{fos03}) of the Lynx arc is reminiscent of
a merging system (several clumps $A$, $B$, $C$ and $D$ joined by a very faint arc), but this is 
misleading since
it is strongly distorted by the effects of the lensing cluster. In their strong lensing
model, \cite{fos03} assume that the Lynx arc consists  of  two different clumps
($A$ and $C$) of the same source,  while 
 $B$ and $D$ are
mirror images of $A$ and $C$ respectively. $A$ and $C$ (or $B$ and $D$) might be two
 components of a merging system. The evidence, however, is not strong enough to confirm this.

\subsection{The continuum problem}

We investigate here whether  the starburst
 luminosity required to excite the strong
emission lines  in the Lynx arc is consistent with the non detection of the stellar
component. 

\cite{fos03}  found a discrepancy (a factor of $\sim$20) 
between the mass of the starburst needed to power the H$\beta$ lumininosity of the Lynx arc and the mass of a
starburst whose continuum would be just marginally detected, in agreement
with the fact that the observed continuum is mostly nebular.
We have done a similar exercise using the SED models that best reproduce
the spectroscopic properties of the Lynx arc.

Using the SEDs by \cite{cer04}
 we have calculated the stellar mass of a burst that can generate
the observed Lynx H$\beta$ luminosity $L(H\beta)$=4.3 $\times$ 10$^{42}$ erg s$^{-1}$
(corrected for magnification, \cite{fos03}). The results are shown in Table 7 for two
of the SEDs  that best reproduce the emission line spectroscopic properties of the Lynx arc (see \S2.2).

The calculations have been done assuming covering factor $CC$=1 (\cite{fos03}). We obtain  M$_{stars} \sim$9 $\times$ 10$^7$ M${\odot}$. 
This is a lower limit, since we have assumed that all ionizing photons
are absorbed by the gas  and there is no dust
reddening.  It is important to note that the uncertainties on the assumed magnification factor
$\mu$ are large. \cite{fos03} found a rather wide range of models that
successfully describe the Lynx arc system. This introduces serious uncertainties
in the estimated cluster mass.

\begin{table*}
\begin{tabular}{llllll}
\hline 
 Metallicity  & Age  & $l$(H$\beta$)  &  Mass$_{H\beta}$ &   F$_{1600}$ \\ 
    $Z_{stars}$/  $Z_{\odot}$         &  Myr  &  erg s$^{-1}$  M$_{\odot}^{-1}$    & M$_{\odot}$  & erg s$^{-1}$ cm$^{-2}$  Hz$^{-1}$  \\ \hline 
 20\%     & 3.4  &  3.2 $\times$ 10$^{34}$ &  9.4 $\times$ 10$^{7}$  & 6.2$\times$10$^{-31}$ \\ 
 40\%     & 3.0  &  3.5 $\times$ 10$^{34}$ &   8.6 $\times$ 10$^{7}$ & 6.4$\times$10$^{-31}$ \\  
 Lynx     &      &                          &                        & 7.9$\times$10$^{-32}$ \\  \hline
 \end{tabular}
\caption{Masses  (Mass$_{H\beta}$) of  instantaneous burst models  required to reproduce 
the H$\beta$ luminosity
of the Lynx arc and expected continuum flux at 1600 \AA\ (rest frame) F$_{1600}$.
The two SEDs are those that best reproduce the emission line spectrum of the Lynx arc (see \S2.2).  $l$(H$\beta$)
is the expected H$\beta$ luminosity per solar mass. The
continuum level expected for an instantaneous  starburst of mass Mass$_{H\beta}$ and SEDs under
consideration
 is $\sim$10 times higher than that  measured for the Lynx arc.} 
\end{table*}

We have then calculated 
 the  stellar continuum at $\sim$1600 \AA ~(rest frame)
emitted by these instantaneous bursts and compared it with the continuum level measured for
the Lynx arc which is $\sim$0.15 $\mu$Jy (observed frame value
 corrected for 
magnification, \cite{fos03}).
This corresponds to  F$_{\nu}$=7.89$\times$10$^{-32}$ erg s$^{-1}$ cm$^{-2}$ Hz$^{-1}$ 
(rest frame). The results are
shown in Table 7.  The theoretical values are a factor of $\sim$10
higher than the  value measured for the Lynx arc. The discrepancy is
 even higher if we take 
 the nebular continuum into account, which according to  \cite{fos03} 
is the dominant continuum component.

As  \cite{fos03}, we therefore conclude that   the stellar  burst that reproduces
 the measured H$\beta$ luminosity
would produce an observable continuum  much ($\sim$10 times) brighter   than  that detected from the Lynx arc. Adding the nebular  contribution, the continuum 
should be even brighter. 
 The conclusion is not affected by uncertainties on the magnification
factor $\mu$ if the continuum and the H$\beta$ flux 
scale in the same way with $\mu$. A similar problem was  reported by \cite{ste00} regarding  
the giant Ly$\alpha$ nebulae associated with Lyman break galaxies. The authors found that the measured UV
 continuum is not enough to explain the high Ly$\alpha$ luminosities of the
nebulae.

We now discuss some possible reasons for this discrepancy in the Lynx arc.

\vspace{0.2cm}

\centerline{\it The AGN scenario}

\vspace{0.2cm}

The previous discussion suggests that the  ionizing source in the Lynx arc is hidden from the observers view.
Such a geometry is characteristic
of type II active galaxies (AGN). \cite{chap04} have recently suggested that the
Lyman ``blobs'' found by \cite{ste00} might be powered by a hidden AGN.
 \cite{bin03} suggested the possibility that  the Lynx arc is a type II active galaxy.
 We, however, strongly support  that  this is a star
forming object for several reasons:

\begin{itemize}
  
\item  an unusual (filtered) ionizing continuum shape is required   to explain the line ratios of the arc
(Binette et al. 2003) within the AGN scenario. Our new models  using normal (i.e. non primordial) stars
show  an excellent agreement with the Lynx arc measured line ratios, 
better than the filtered AGN models (see for instance, [OIII]$\lambda$5007/Hb, NV/Hb,  CIV/H$\beta$,
CIII]/H$\beta$ in Table 1 of \cite{bin03}). I.e. there is no need for an exotic AGN
model to reproduce the line ratios.

\item  the very narrow lines of the Lynx arc (FWHM of non resonant lines $<$100 km s$^{-1}$, \cite{fos03})
 are typical of star forming objects. AGNs usually show  line
widths of  few hundred km  s$^{-1}$. Although there are examples of extended emission line
regions in active galaxies with narrower lines, these cases are exceptional.

\item  The strongest discrepancies of the filtered AGN model disappear with our new models:
these are the severe underprediction of the SiIII] lines and  the too high
electron temperature ($\sim$20 000 K) predicted by the filtered AGN models.

\end{itemize}

We therefore reject the possibility that the Lynx arc is an active galaxy.

\vspace{0.2cm}

 \centerline{\it A hidden starburst}

\vspace{0.2cm}

 Hidden starbursts have  
 been proposed to explain the spectroscopic properties  of
 the metal poor blue compact galaxy HS 0837+4717
(Pustilnik et al. 2004). In this object, the extinction of the narrow emission lines
from the giant HII region is low, but the large Balmer decrement ($\sim$19$\pm$4)
of the broad
components suggests that part of the current starburst is highly obscured
by dust. A highly obscured starburst has also been revealed by mid-IR observations
of the very low metallicity galaxy SBS 0335-052 ($Z$=1/41 $Z_{\odot},~ $\cite{hunt01}). 
Roughly 3/4 of the star formation in this object occurs within the obscured cluster.

It is not clear however, that this scenario solves the  the Lynx arc problem.
In  HS 0837+4717 and
SBS 0335-052 the non-obscured regions
emit both strong continuum and emission lines.  I.e., if optical emission lines 
are detected, the ionizing stars are also detected. This is not the case for the Lynx arc.

\vspace{0.2cm}

\centerline{\it Differential gravitational lensing amplification}
\vspace{0.2cm}

The different spatial distribution between ionized gas and young stars can be the
 solution to the Lynx  continuum
problem. 

The discrepancy
between reddening values derived using the UV continuum and 
 the ratios of the Balmer lines  found in some  star forming galaxies (e.g. \cite{mas98})
has been explained as due to 
 ionized gas and stars having different spatial distributions and suffering different extinction by dust.
 When this discrepancy is found, the UV continuum seems to be less
affected by extinction than the emission lines. The opposite case would
apply to the Lynx arc. 
Spatial decoupling between ionized gas and young stars has been
observed in some star forming galaxies 
(e.g. \cite{ape98}) and proposed by several authors to explain different issues
(e.g. \cite{mas98}, \cite{rosa00})

The degree of amplification suffered by the source  undergoing lensing depends strongly 
on both the position close to the critical lines
and the geometry of the lensed object, so that   separate regions  can suffer different degrees of
amplification. If stars and gas have different spatial distributions in the Lynx arc,
 emission lines and stellar continuum   can suffer different
 magnification due to the intervening cluster. Differential amplification between
lines and continuum has been observed in several quasars (e.g. \cite{rac92}).
Detailed gravitational lensing models would be necessary to verify this possibility.

\vspace{0.2cm}

\centerline{\it A primordial stellar population}
\vspace{0.2cm}

The SED of metal-free  stars is characterized by effective temperatures on the Main Sequence around
10$^5$ K, hotter 
 than their counterparts
of equal mass but finite metallicity (e.g. \cite{tum00}). The UV recombination emission lines, such as 
HeII$\lambda$1640, 
 would have
extreme equivalent widths more than an order of magnitude larger than the expectation 
for a normal cluster of hot stars with the same total mass and a Salpeter IMF. Bromm, Kudritzki \& Loeb
(2001) estimated that a cluster of 10$^6$ M$_{\odot}$ of popIII stars with a heavy IMF would produce
$\sim$2.3$\times$10$^{42}$ erg s$^{-1}$ in the HeII$\lambda$1640 line and
a spectral luminosity per unit frequency of 1.8$\times$10$^{27}$ erg s$^{-1}$ Hz$^{-1}$.
For the observed luminosity
of the HeII$\lambda$1640 in the Lynx arc ($\sim$4.4$\times$10$^{42}$ erg s$^{-1}$) a cluster 
of $\sim$2$\times$10$^6$ M$_{\odot}$ of primordial stars would be required.  Such a cluster  would
produce a spectral flux per unit frequency of 3.1$\times$10$^{-32}$ erg s$^{-1}$ cm$^{-2}$ Hz$^{-1}$
at the distance of the Lynx arc. This is a factor of $\sim$2.3  below the continuum measured 
for the Lynx arc. The discrepancy, therefore, disappears and this result 
is consistent with the conclusion by \cite{fos03} that the continuum is mostly nebular.

The possibility that the Lynx arc is a popIII  object was proposed and discussed in detail
by \cite{fos03}. It is difficult to explain, however, nebular abundances as high as $\sim$10\%
if the stellar population is primordial.

\section{Summary and conclusions}

We have characterized the physical properties (electron temperature, density, chemical
abundances) of the ionized gas and the ionizing stellar  population in the Lynx arc, a gravitationally
amplified HII
galaxy  at $z$=3.36. 
 The temperature sensitive ratio [OIII]$\lambda\lambda$1661,1666/$\lambda$5007
implies an electron temperature 
$T_e$=17300$^{+500}_{-700}$ K, in good agreement within the errors with photoionization model predictions.
 The UV doublets 
imply the
existence of a density gradient in this object, with a highly ionized high density 
region (0.1-1.0 $\times$ 10$^5$ cm$^{-3}$) and a low density region
($<$3200  cm$^{-3}$) with lower ionization state.

Both the photoionization modeling and standard techniques of chemical
abundance determination  imply that the gas metallicity is 
$\sim$10$\pm$3\% Z$_{\odot}$. Both methods suggest that  
nitrogen is overabundant with  N/O$\sim$2.0-3$\times$[N/O]$_{\odot}$, unless a density gradient
produces this apparent effect.
We do not find evidence for
Si overabundance as Fosbury et al. (2003). The reason is the different shape of the ionizing
continuum assumed by these authors (80 000 K black body, much hotter than the dominant ionizing
stars in our models). 

Photoionization models imply that the ionizing
stars have very young ages $\la$5 Myr. Since the emission lines trace the properties of 
the present burst only, nothing can be said
about the possible presence of an underlying old stellar population. 
 Instantaneous burst models with Z$_{star}$$>$5\% Z$_{\odot}$ and
ages $\sim$2.8-3.4 Myr (depending on Z$_{star}$), are in excellent agreement 
with the
Lynx spectrum, including the strong  He$^{+2}$ emission.  At this age
Wolf Rayet stars make an important contribution to the hard ionizing
luminosity and they are responsible for the excitation of the He$^{+2}$ emission.
In such case, we infer the existence of
$\sim$2.5$\times$10$^4$ WR stars in the Lynx arc.
Alternative excitation mechanisms for He$^{+2}$, however, cannot be  discarded.

Therefore, the Lynx arc is  a low metallicity HII galaxy  that
is undergoing a  burst of star formation of $\la$5 Myr  age. One possible scenario that
 explains the emission line spectrum of the Lynx arc, the strength of the nitrogen lines
   and the strong
He$^{+2}$ emission is that the object has experienced a merger event that has
triggered a   powerful starburst phase. Wolf Rayet stars have been formed and
contribute to a fast chemical enrichment of the interstellar medium.

As  Fosbury et al. (2003), we find a factor of $>$10 discrepancy   between the mass of the instantaneous burst 
implied by the luminosity of the H$\beta$ line and the mass implied by the continuum level
measured for  the Lynx arc. 
 We have discussed several possible solutions to this problem.
The most satisfactory explanation  is that gas and stars have different spatial distribution
so that the emission lines and the stellar continuum suffer different gravitational 
amplification by the intervening cluster. Detailed gravitional lensing models are needed to test
the vailidity of this scenario.

\section*{Acknowledgments}
We thank an  anonymous referee for providing very useful comments that helped to improve this
paper substantially.
We thank Valentina Luridiana for useful scientific discussions and 
        Andrew Humphrey for providing Figure  1. M. Villar-Mart\'\i n and M. Cervi\~no are supported
by the Spanish National program Ram\'on y Cajal. We acknowledge support by the Spanish Ministry of
Science and Technology (MCyT) through grant AYA-2001-3939-C02-01.

\end{document}